\documentclass[%
aps,
jmp,%
amsmath,amssymb,
reprint,%
]{revtex4-2}
\usepackage{graphicx}% Include figure files
\usepackage{subfigure}
\usepackage{dcolumn}% Align table columns on decimal point
\usepackage{bm}% bold math
\usepackage[mathlines]{lineno}% Enable numbering of text and display math
\usepackage{graphicx}
\usepackage{graphicx}% Include figure files
\usepackage{subfigure}
\usepackage{dcolumn}% Align table columns on decimal point
\usepackage{bm}% bold math
\usepackage[mathlines]{lineno}% Enable numbering of text and display math
\graphicspath{{figure/}}
\usepackage{array}
\usepackage{braket}
\usepackage{diagbox}
\usepackage{amsmath}
\usepackage{appendix}
\usepackage{multirow}
\usepackage{bm,color,bbm}

\newcommand{\blk}{\color{black}}

\usepackage{float}
\newcolumntype{.}{D{.}{.}{-1}}
\usepackage{amsopn}
\usepackage[colorlinks,linkcolor=blue, anchorcolor=blue, urlcolor=blue, citecolor=blue]{hyperref}
\usepackage{epstopdf}

\begin{document}

	\title {A cost-efficient quantum access network with qubit-based synchronization}% Force line breaks with \\
	\author{Chunfeng Huang$^{1}$, Ye Chen$^{1}$, Tingting Luo$^{1}$, Wenjie He$^{1}$, Xin Liu$^{1}$, Zhenrong Zhang$^{2}$, and Kejin Wei$^{1,*}$}
	\address{
		$^1$Guangxi Key Laboratory for Relativistic Astrophysics, School of Physical Science and Technology,
		Guangxi University, Nanning 530004, China\\		$^2$Guangxi Key Laboratory of Multimedia Communications and Network Technology, School of Computer Electronics and Information, Guangxi University, Nanning 530004, China\\
		$^*$Corresponding author: kjwei@gxu.edu.cn
	}

		\begin{abstract}
		
		Quantum Key Distribution (QKD) is a physical layer encryption technique that enables two distant parties to exchange secure keys with information-theoretic security. In the last two decades, QKD has transitioned from laboratory research to real-world applications, including multi-user quantum access networks (QANs). This network structure allows users to share a single-photon detector at a network node through time-division multiplexing, thereby significantly reducing the network cost. However, current QAN implementations require additional hardware for auxiliary tasks such as time synchronization. To address this issue, we propose a cost-efficient QAN that uses qubit-based synchronization. In this approach, the transmitted qubits facilitate time synchronization, eliminating the need for additional synchronization hardware. We tested our scheme by implementing a network for two users and successfully achieved average secure key rates of $53.84$ kbps and $71.90$ kbps for each user over a 50-km commercial fiber spool. In addition, we investigated the capacity of the access network under cross-talk and loss conditions. The simulation results demonstrate that this scheme can support a QAN with 64 users with key rates up to 1070~bps. Our work provides a feasible and cost-effective way to implement a multi-user QKD network, further promoting the widespread application of QKD.

    	\end{abstract}
	    \maketitle
	    
    	\section{Introduction}
        The security of information exchange is of great significance to human society. Quantum key distribution (QKD) allows two authorized users to exchange keys with information-theoretical security, and the security of the key bits is guaranteed by the laws of quantum physics~\cite{1984BEN,1999Lo,2000Shor,2004Gottesman}. Since Bennett and Brassard proposed the first QKD scheme~\cite{1984BEN}, known as the BB84 protocol, QKD has rapidly matured through various theoretical schemes~\cite{2005Lo,2005Wang,2012Lo,2018lucamarini,2018Ma,2018Wang,2022zeng-mode} and continuous technological innovations~\cite{2018Boaron,2020grunenfelder,2020Wei,2021paraiso,2023zhou,2023grunenfelder-fast,2023li,2023liu,2023chen-continuous}. Recently, the secure key rate of QKD implementations has reached several dozen megahertz~\cite{2023grunenfelder-fast,2023li}, and the record-breaking distance has been extended to 1000 km~\cite{2023liu}. Many multi-user QKD networks have been reported, including metropolitan networks~\cite{2010Chen,2011Sasaki,2018Bunandar,2021avesani,2021chen-implementation}, integrated space-to-ground quantum communication networks~\cite{2021Chen}, and networks with untrusted relays~\cite{2022Zhong,2022Fan-Yuan}. 
 
        The quantum access network (QAN) is a well-known QKD network structure that provides last-mile services for multiple users to access the QKD infrastructure~\cite{2010Choi,2013.Frohlich.Nature,2015.Frohlich.Scientificreports,liu2022sync,2023WangCV,sun2018,Wang2021,Huang2021}. In general, there are two configurations of an access network. The first, termed a ``downstream network’’, involves placing the QKD transmitter on the network node, and each user has a QKD receiver~\cite{sun2018,Wang2021,Huang2021}. The second configuration, known as the ``upstream network’’, involves each user having a transmitter, and the receiver is located at a network node~\cite{2010Choi,2013.Frohlich.Nature,2015.Frohlich.Scientificreports,liu2022sync,2023WangCV}. In a downstream access network, the cost of a single-photon detector is usually beyond users' affordability. In addition, detectors with complex designs are the most frequently attacked devices~\cite{2011Sauge,2018Qian-Hack,2019Wei}, requiring further consideration of detection security in the network. In contrast, users of the upstream access network require only a transmitter, and the detectors are provided by the node, which greatly reduces the network cost~\cite{2013.Frohlich.Nature}. Furthermore, upgrading the security of the network is easier in a network structure that is independent of the measurement device~\cite{2016Tang,2017Yin,2019Wang,2021Fan-Yuan,2022Park}. Owing to their cost advantages and upgrade flexibility, QANs have gained widespread attention and have been validated in a metropolitan network~\cite{2019dynes}.
 
        Even so, in existing upstream networks, each transmitter not only must possess quantum state preparation equipment but also requires additional optical synchronization devices to ensure that photons fall into predetermined time slots. This makes the transmitter equipment more complex. Furthermore, previous studies  have shown that in co-fiber transmission schemes, the use of additional optical synchronization devices introduces Raman scattering, which severely impairs the transmission of quantum signals~\cite{2010Choi,2015.Frohlich.Scientificreports}. Therefore, it is necessary to add additional devices, such as filters, to improve the performance of quantum signal transmission. This further increases construction costs and network complexity.
 
        In this study, we propose and demonstrate a cost-efficient QAN scheme that exploits a recently reported qubit-based frame synchronization method~\cite{2023Chen.qubitbased}. In our scheme, we implement time synchronization between each transmitter and the node by relying on the flying qubits, thereby eliminating the need for additional synchronization hardware. This approach further reduces the network cost because each transmitter requires only devices for qubit generation. To verify the feasibility of the network scheme, we implemented a polarization-coding QAN with two users. Experimental results show that the two users on the network achieved secure key rates of $53.84$ kbps and $71.90$ kbps, respectively, over $56$-km commercial fiber spool. Additionally, we analyze the capacity of the access network in the presence of cross-talk and loss using the parameters of the cross-talk noise measured in our setup. Simulation results show that the network can support up to $64$ users with key rates up to 1070~bps. Our work provides a feasible and cost-effective way to implement a multi-user QKD network, further promoting the widespread application of QKD.
	
    	\section{Qubit-based access network architecture}

    	\subsection{Network architecture}
\begin{figure}[htp]
	    	\centering
	    	\subfigure{
	    		\label{fig_withSync}
	    		\includegraphics[width=0.80\linewidth]{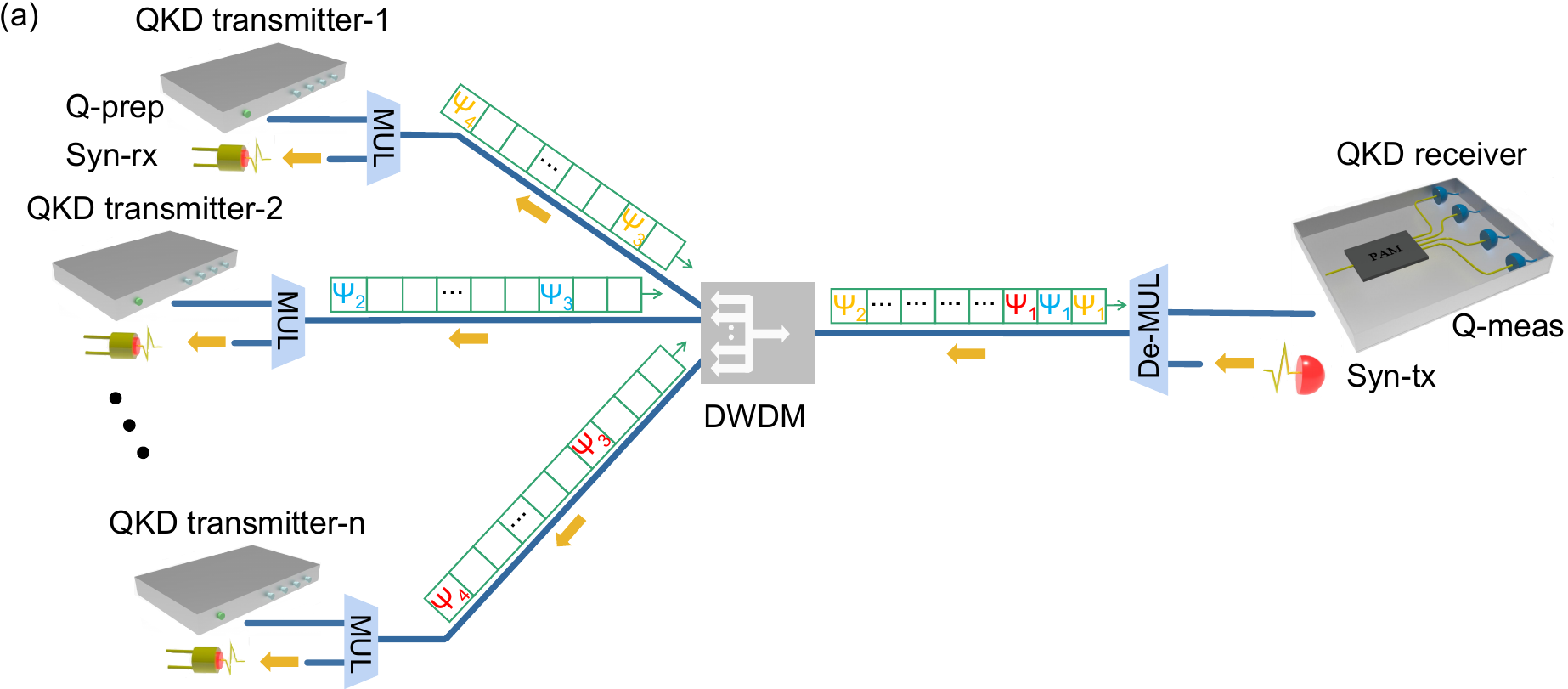}
	    	}
	    	\\
	    	\centering
	    	\subfigure{
	    		\label{fig_withoutSync}
	    		\includegraphics[width=0.80\linewidth]{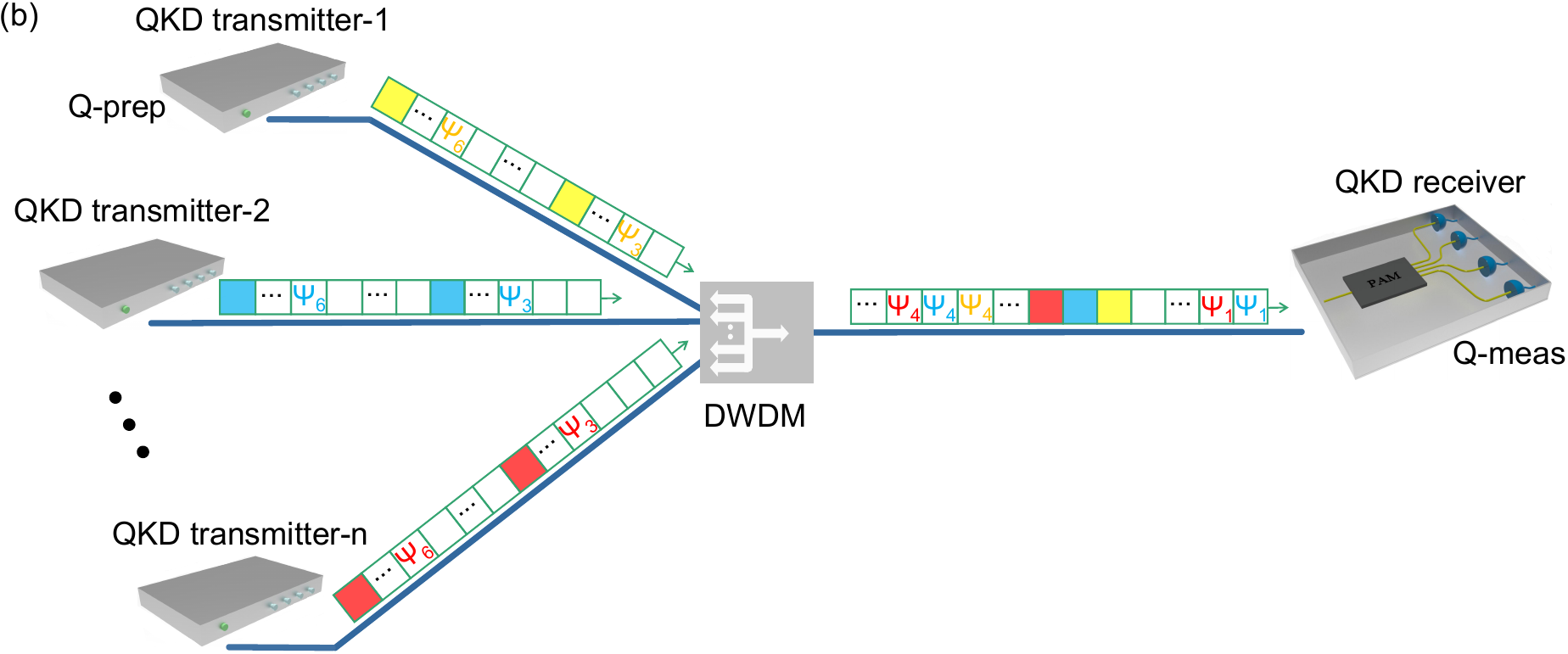}
	    	}
	    	\centering
	    	\caption{Each QKD transmitter prepares and sends quantum signals in different time slots, which are transmitted to the QKD receiver using a dense wavelength division multiplexer (DWDM). (a) A typical quantum access network architecture. The orange arrows represent classical synchronization optical signals. They are coupled with the quantum signals through a multiplexing module (MUL). (b) Our access network scheme. The colored-filled positions include the quantum signals used for time synchronization.}
	    \end{figure}

	    First, we briefly introduce a typical QAN. As shown in Figure~\ref{fig_withSync}, each user has a transmitter, where some devices are used to prepare quantum signals, and an additional synchronization system is used to achieve time synchronization with the network node. The signals emitted by each user are coupled through dense wavelength-division multiplexers, combined in time-multiplexed pulse by pulse and sent to the network node. With the help of synchronization devices for each user, quantum signals can accurately fall into pre-set time slots, ensuring that the QKD receivers in the node can accurately detect the quantum signals from each user. 
	    
        Figure~\ref{fig_withoutSync} shows the proposed QAN scheme. Our architecture is identical to that of a traditional QAN, except that we removed the synchronization devices and retained only the devices for preparing quantum states. Furthermore, this scheme requires a passive basis decoder and free-running single photon detectors. To achieve an accurate key distribution with the network node, in addition to sending completely random BB84 quantum signals, as in traditional network methods, we also periodically insert publicly known synchronization quantum signals (represented by yellow, blue, or red squares in the figure). These states are then coupled together and sent to the network node. The network node uses a receiver to detect and analyze these states, recording their arrival times using its local clock. The data are then processed using synchronization algorithms that enable the generation of raw secure keys for each user.
        
        \begin{figure*}[htp]
        	\centering	
        	\includegraphics[width=0.8\linewidth]{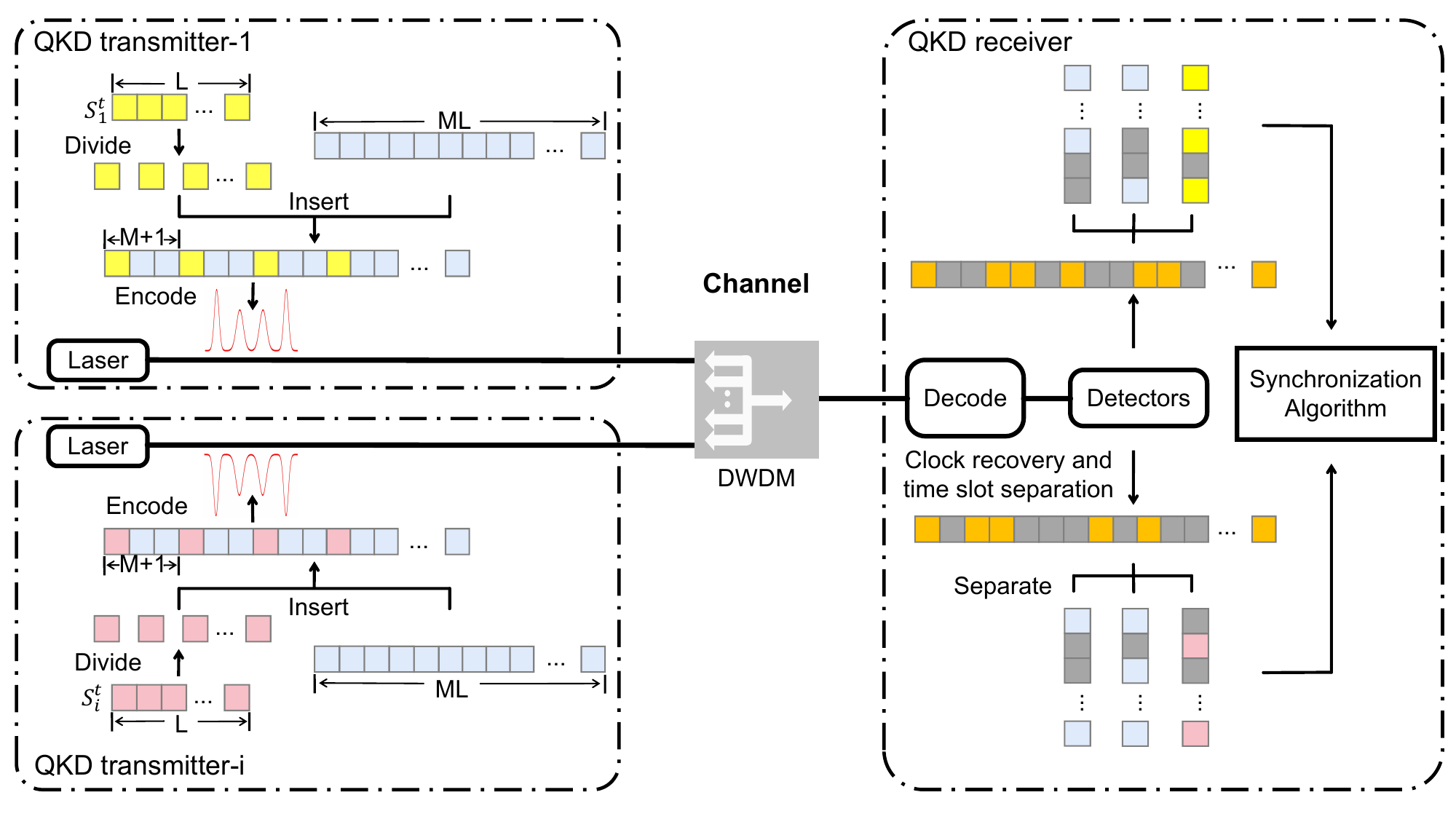}			
        	\caption{Schematic of the operation of access network using a qubit-based synchronization method. Blue squares denote random bits. Yellow and pink squares represent synchronization bits for distinct transmitters. The synchronization string $S_{1}^{t}$ is broken into individual bits and uniformly inserted at intervals of $M$ into the random bit string, forming a bit frame of length $(M+1)\times L$. The received frame contains detected bits (orange squares) and undetected bits (gray squares). Every received frame must include a synchronization bit string and $M$ random bit strings.}
        	\label{fig_withoutSync_operation}
        \end{figure*} 
        
        Figure~\ref{fig_withoutSync_operation} illustrates the entire operation of our network using qubit-based synchronization, which can be divided into the following steps:
        
       1) Each user first utilizes a laser controlled by a local crystal oscillator to emit weak light pulse to the receiver. The receiver utilizes a free-running single-photon detector to measure its arrival time, ensuring synchronization between each user and the receiver for a specific duration. Subsequently each user implement time-division multiplexing based on this synchronization.               
        
       2) Subsequently, each user  generates a unique publicly known synchronization string of length $L$, and then generates a string of completely random bits used to modulate the four states of BB84. The synchronization string generated by each user can be regarded as the user's identifier. The user then divides this string into $L$ individual bits and periodically inserts these into the bits used for random modulation to construct a bit frame. The frame is used to encode a sequence of quantum signals based on the transmitter's local clock.  This process ensures that only qubits from one transmitter reach the receiver simultaneously. Qubits from different transmitters are coupled and sent using time-division over a lossy channel.  
        
      3) The receiver measures the quantum signals from different transmitters and analyzes their states, recording time-of-arrival measurements using its local clock. The receiver's main goal is to determine the sources of the qubits within the detection time slots and identify the positions of the received bits in the transmitter's random bit string. These operations are critical for the accurate generation of raw key bits between each transmitter and receiver and for effective sifting of key bits. To achieve this, the receiver must accomplish three tasks: $i$) Recover the clock $\tau^{R}$ from the detection events. $ii$) Separate the qubits from different detection time slots. $iii$) Derive the corresponding transmitters for the time slots and calculate the time delay between the received and sent strings. Below, we explain how to use the transmitted quantum states to accomplish these three tasks, enabling QANs without the need for additional synchronization hardware.   	
        
        \subsection{User identification and time recovery using qubit states}
        Here we describe the construction of the transmitted bit frames. For example, transmitter-$1$ generates a synchronization bit string $S_{1}^{t}$ of length $L$ and a random bit string of length $M \times L$. The string $S_{1}^{t}$ is decomposed into individual bits and uniformly inserted with a preset bit interval $M$ into the random bit string, constructing a bit frame of length $(M+1) \times L$, as shown in Figure~\ref{fig_withoutSync_operation}. Similar to previously described methods~\cite{2023Chen.qubitbased,2020Calderaro}, $S_{1}^{t}$ exhibits periodic correlations and is publicly disclosed to the receiver, whereas the random bit string is used to generate secure keys and must be kept secret. Similarly, the synchronization strings are encoded on two orthogonal bases; for example, the horizontal and vertical polarizations of the $z$-basis are assigned as $+1$ and $-1$. At the same time, there is no correlation between $S_{1}^{t}$ and the synchronization string $S_{i}^{t}$ of transmitter-$i$, enabling the receiver to identify the source of qubits through the specific string.
        
        The receiver measures quantum signals from different transmitters and records their arrival times $t_{a}^m$ using its local clock. Owing to channel losses and detection noise, the receiver cannot determine the delay of the detection time slot and thus cannot identify the source of the quantum signals. Subsequently, we describe how the receiver achieves time recovery based on the measurements and autocorrelation properties of the synchronization bit string.
        
        Our method first recovers the clock $\tau^{R}$, enabling the receiver to accurately reconstruct the bit intervals in the raw key between consecutive detections. The receiver initially estimates a clock $\tau_{0}^R$ from the arrival time based on the Fourier transform, and then refines it using the least-squares algorithm to obtain an accurate value. To separate the detection time slots, the receiver employs the recovered clock $\tau^{R}$ to calculate the modulo of the quantum signal's arrival time. This procedure effectively segregates qubits into distinct time slots. Nevertheless, owing to channel losses and detection noise, the receiver faces difficulty in determining the source of qubits accurately within these time slots. In other words, the demarcated slots may deviate from the slots pre-designated by the calibration process. Consequently, the receiver must identify the transmitters associated with each time slot before calculating the time offset between the measured string and the transmitter's raw string.
        
        Based on the above, the receiver can extract a received frame from each time slot. Guided by the transmitted frame structure of periodically inserted synchronization bits, the receiver translates the received frame into a synchronization qubit string of length $L$, and $M$ sets of random qubit strings each of length $L$. Considering that the send synchronization string is encoded only on the $z$-basis, the receiver similarly assigns the values $+1$ and $-1$ to the two orthogonal states in the $z$-basis, whereas received qubits that are either undetected or collapsed into the $x$-basis are assigned a value of $0$. The receiver then associates the strings from different time slots with $S_{1}^{t}$. Exploiting the autocorrelation property of the synchronization string $S_{1}^{t}$, as the maximum cross-correlation value emerges, the receiver can determine the corresponding time slot for transmitter-$1$ and associate received bits with the bits in the transmitter's raw string. A detailed description of the synchronization method is given in Appendix~\ref{Appendix_A}. Furthermore, we discuss some practical issues for the synchronization in Appendix~\ref{Appendix_B}.
	
     	\section{Experiments}
     	
     	\begin{figure}[htp]
     		\centering
     		\includegraphics[width=0.80\linewidth]{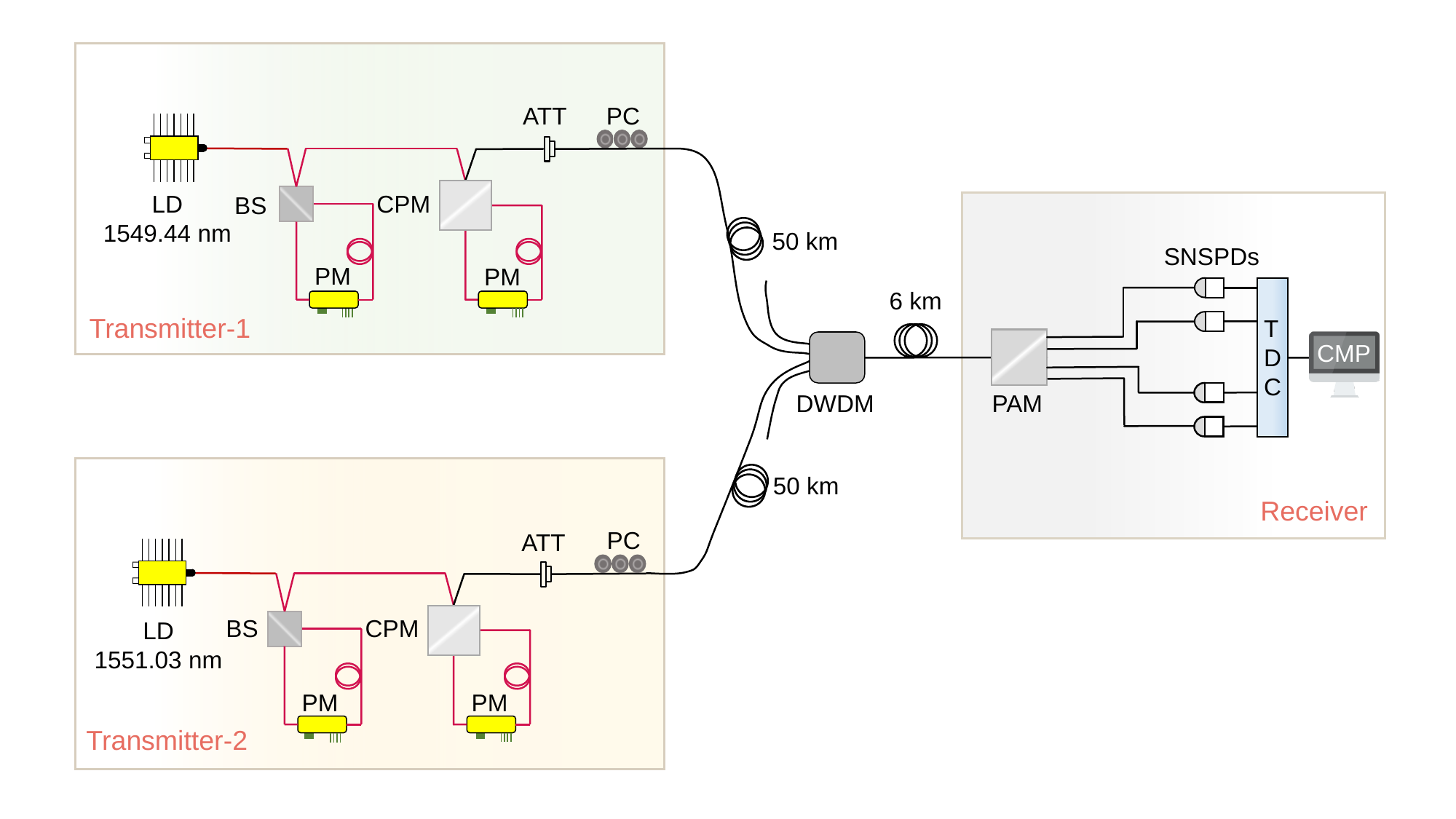}
     		\caption{Schematic diagram of our experimental setup for the upstream quantum access network. LD:  commercial laser diode; BS: beam splitter of 75:25; PM: phase modulator; CPM: customized polarization module; ATT: optical attenuator; PC: polarization controller; DWDM: dense wavelength division multiplexing; PAM: polarization analysis module; SNSPDs: superconducting nanowire single photon detectors; TDC: time-to-digital converter. CMP: a computer performing time synchronization algorithm.}
     		\label{fig_setup}
     	\end{figure}
    	To validate the feasibility of the proposed scheme, we conducted a QAN experiment with two users. The experimental setup of our network is illustrated in Figure~\ref{fig_setup}. Each user adopts a polarization-coding QKD transmitter~\cite{Ma2021} running the decoy-state BB84 protocol. The transmitter uses a commercial laser (LD, WT-LD200-DL, Qasky Co., Ltd.) to generate light pulses with a pulse width of $200$ ps and a repetition rate of $50$ MHz. The center wavelengths of the light pulses are $1549.44$ nm and $1551.03$ nm. The light pulses are fed into a Sagnac-based intensity modulator, which produces signal and decoy pulses with a fixed intensity ratio. The intensity modulator contains a fixed $75:25$ beam splitter (BS) and a phase modulator (PM), achieving a fixed $4:1$ intensity ratio between the decoy state and the signal state. The ratio is closed optimum parameter for all distance. Then, the light pulse is coupled into a Sagnac-based polarization modulator which comprises a customized polarization module (CPM) and PM to prepare four BB84 polarization states $|\psi\rangle=( |H\rangle+e^{i \theta}|V\rangle ) / \sqrt{2}, \theta \in\{0, \pi / 2, \pi, 3 \pi / 2\}$, where $\theta \in\{0, \pi\}$ $(\theta \in\{\pi / 2, 3 \pi / 2\} )$ represents the $z(x)$-basis. Subsequently, the light pulses enter an optical attenuator (ATT) to be attenuated to the single-photon level and are sent to the receiver over a fiber spool with a length of $50$ km. 
    	
    	The quantum signals pass through a lossy channel to reach the receiver, which includes a customized passive polarization analysis module (PAM). This module serves as a polarization reference for multiple transmitters. The coupling ratio of the two measurement basis of the PAM is set to $9:1$. Each user uses a polarization controller (PC) to calibrate their reference frame independently with respect to the receiver. Two users are connected to a $100$G $1 \times 4$ dense wavelength division multiplexer (DWDM) via a $50$ km fiber. We add a $6$ km optical fiber between the DWDM and the network node, similar to the QAN of Fr\"{o}hlich et al.~\cite{2013.Frohlich.Nature}, where the passive optical component and receiver are not co-located. The QKD receiver, located at the network node, includes four superconducting nanowire single-photon detectors (SNSPDs, Photon Technology Co., Ltd.) with $50\%$ detection efficiency and $50$ Hz dark count. Photon detection events are recorded using a high-speed time-to-digital converter (TDC, Timetagger20, Swabian Instruments). The detection data are input into the time synchronization algorithm, which is executed on a computer.
    	
    	\section{Results}
    	
    	\begin{figure*}[htp]
    		\centering
    		\includegraphics[width=0.85\linewidth]{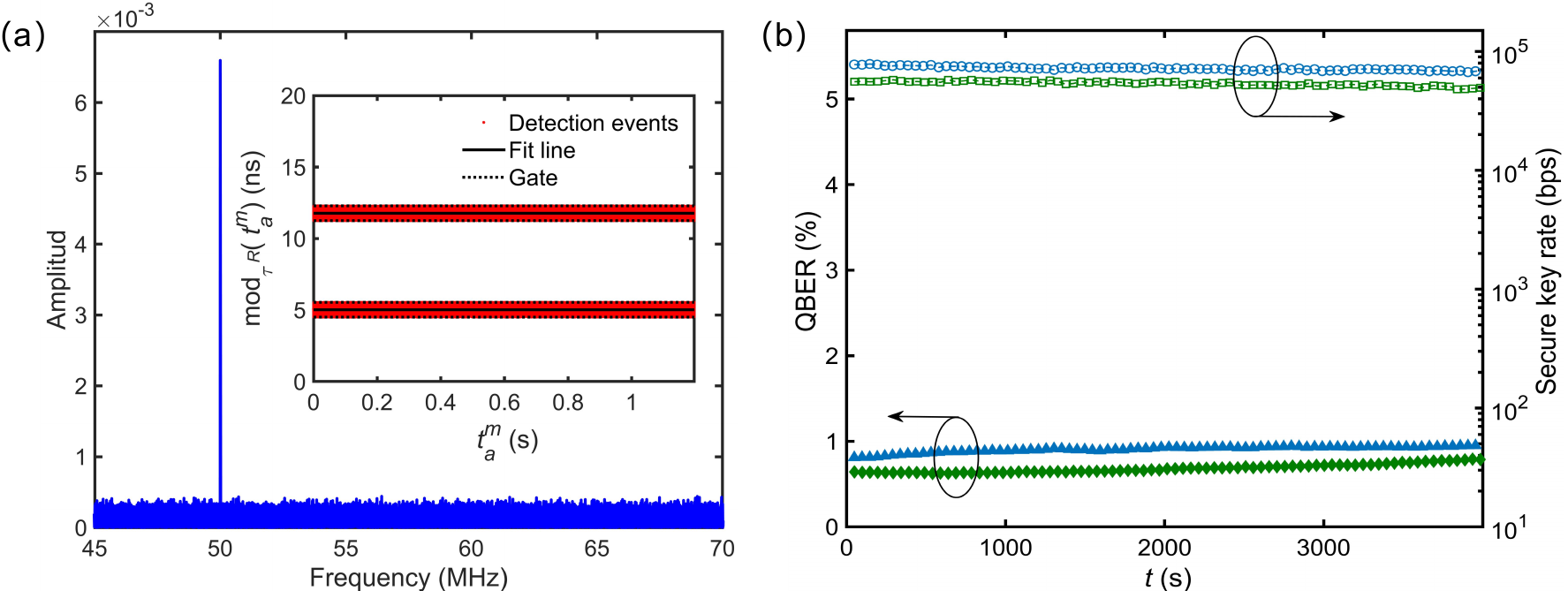}
    		\caption{Experimental results in a quantum access network with two transmitters using our qubit-based method. (a) The intermediate results of the synchronization algorithm. The Fourier transform estimates the clock $\tau_{0}^{R}$. The inset indicates the distribution of detection events separated from different time slots after accurate clock recovery. Red dots represent detection events. The solid black line is a line fitted based on the distribution of user detection events. The distance between the black dashed lines for each user is the detection gate width based on the fit line, which allows filtering some of the noise. (b) The quantum bit error rates (QBERs) and secure key rates for each user.  The QBER (filled green diamonds, transmitter-1; filled blue triangles, transmitter-2) and secure key rate (open green squares, transmitter-1; open blue circles, transmitter-2) for each $z$ basis sifted key block with a size of $10^{7}$.}
    		\label{fig_results}
    	\end{figure*}
    
        \begin{figure*}[htp]
    	\centering
        \includegraphics[width=0.85\linewidth]{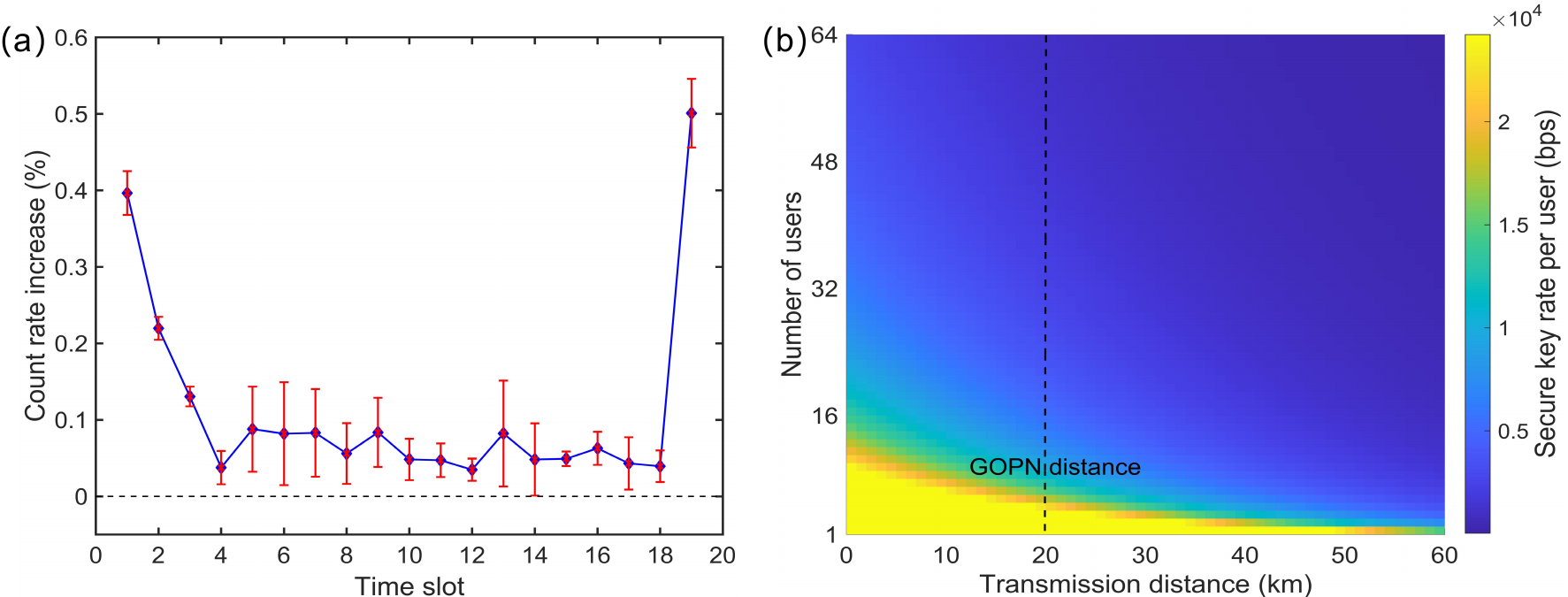}
    	\caption{(a) The relative increase in count rate caused by cross-talk. The cross-talk caused by the adjacent time slot are $0.43\%$ and $0.55\%$, respectively. The error bars correspond to one standard deviation of  measurements. (b) The secure key rate of a single user in a $64$-user quantum access network with different number of active users.}
    	\label{fig_crosstalk_simulation}
        \end{figure*}
        We implemented a QAN with qubit-based synchronization accommodating two users over commercial optical fibers. The channel transmission losses of transmitter-1 and transmitter-2 were $12.164$ dB and $12.131$ dB, respectively. We numerically optimized the secure key rate for the channel transmittance of these links by selecting appropriate parameters. Owing to the similar channel transmittance for both users, their parameters were set as follows: the intensities of the signal and decoy states were $\mu=0.52$ and $\nu=0.13$, respectively. The probabilities of preparing the signal and decoy states were $P_{\mu}=0.69$ and $P_{\nu}=0.31$, respectively. Additionally, the probability of preparing the state in the $z$-basis was $0.9$, and that in the $x$-basis was $0.1$. The bit frame used to modulate the quantum states had a length of $2\times10^5$, and the synchronization string had a length of $L=10^5$. Thus, the ratio $M$ is equal to $1$. The lower bound of the secure key rate is given by 
    	
    	\begin{equation}
    	\begin{aligned}
    	R \geqslant & (s_{z, 0}^{L}+s_{z, 1}^{L}(1-h\left(e_{z,1}^{ph}\right))-\lambda_{EC}\\ & -6 \log _{2} \frac{19}{\varepsilon_{sec}}-\log _{2} \frac{2}{\varepsilon_{cor}})\cdot q \cdot f /N,
    	\end{aligned}
    	\label{lfinite}
    	\end{equation}
    	where $s_{z,0}^{L}$ represents the lower bound of vacuum events in the $z$-basis, $s_{z, 1}^{L}$ represents the lower bound of single-photon events in the $z$-basis, and $e_{z,1}^{ph}$ is the phase error rate of single-photon events. $\lambda_{EC}=n_{z}f_{e}h(e_{z})$ represents the number of announced bits in the error-correction stage, where $e_{z}$ denotes the quantum bit error rate (QBER), and $f_{e}=1.16$ is the error correction factor. $h(x)$ represents the binary Shannon entropy, denoted as $h(x)=-x \log _{2}(x)-(1-x) \log _{2}(1-x)$. $\varepsilon_{sec}$ and $\varepsilon_{cor}$ are the secrecy and correctness criteria, equal to $10^{-9}$ and $10^{-15}$, respectively. $f$ is the repetition frequency and $N$ represents the total number of optical pulses, including synchronization and random pulses. $q=M/(M+1)$ denotes the ratio of signals for key distribution.
    	
    	We conducted time recovery on approximately $1$ s of detection data per iteration. The intermediate results of the synchronization algorithm are depicted in Figure~\ref{fig_results}(a). The Fourier transform estimates the clock $\tau_{0}^{R}$ of the detected quantum signals from different users to be $50$ MHz. The inset shows the distribution of the detected events of the two users obtained through processing, after recovering the precise clock $\tau^{R}$.
    	
    	For the finite-size analysis, we consider sifted keys with a size of $n_z=10^7$, accumulating approximately $50$ s of raw key data. As shown in Figure~\ref{fig_results}(b), the average QBER for transmitter-1 (transmitter-2) was $0.69 \% ~(0.91\%)$. Here,  transmitter-2 exhibits a higher QBER than transmitter-1 due to a larger inherent optical misalignment. Average secure key rates for transmitter-1 and transmitter-2 were $53.84$ kbps and $71.90$ kbps, respectively.
    	
    	To estimate the capacity of our scheme, we conducted a detailed analysis of the network's performance with an increasing number of users, considering our system parameters. As the number of users increased, cross-talk between users and increased loss lead to a reduction in the signal-to-noise ratio of the received frame. These effects negatively affect network time synchronization and operation. Similar to the analysis conducted by Fr\"{o}hlich et al.~\cite{2013.Frohlich.Nature}, we assessed the performance of each transmitter's secure key rate in the presence of cross-talk and loss. Initially, we evaluated the cross-talk introduced for the time slot of transmitter-1 by measuring the excess counts introduced from different time slots. The cross-talk for transmitter-1 was determined by recording the increasing counts both with and without the presence of signals from transmitter-2.
    	
    	Based on the distribution of detection events in the clock window, we set the width of the detection time slot to $1$ ns. For different transmitters with a repetition rate of $50$ MHz, this allows up to $20$ users to access the network simultaneously. Figure~\ref{fig_crosstalk_simulation}(a) shows the relative increase in count rate introduced by transmitter-2 at different time slots. It can be clearly seen that time slots in closer proximity to transmitter-1, specifically time slots 1-3 and 19, display a more rapid count increase and exhibit increased crosstalk. Because there are statistical fluctuations in transmitter-1's signal, the effect of cross-talk due to the remaining time slots was not considered.
    	
    	The performance of users in an access network can be simulated based on cross-talk test data. Figure~\ref{fig_crosstalk_simulation}(b) shows the secure key rate of a single user in a $64$-user network with varying numbers of active users and channel loss. The channel loss includes losses from the fiber spool and a $1\times64$ splitter, which has a splitting ratio of $19.5$ dB, as obtained from a typical QAN experiment~\cite{2013.Frohlich.Nature}. Here, since splitters typically exhibits greater loss compared to DWDM. Therefore, we utilize it to model a worst-case scenario when estimating the QAN capacity. 
    	
    	For a small number of users, the transmission distance can reach $60$ km, enabling each user to achieve a secure key rate of approximately $10^{4}$ bps. If the transmission distance is limited to the maximum distance for a gigabit passive optical network (GPON) of $20$ km, the network scheme can accommodate 64 users simultaneously. Under these conditions, the overall transmittance was $27.6$ dB, and each user could achieve a secure key rate of $1070$ bps. The simulation, which considers cross-talk, is detailed in Appendix~\ref{Appendix_C}.
    	
     	\section{Conclusion and Discussion}
        We proposed and experimentally demonstrated a cost-efficient QAN that eliminates the need for additional synchronization hardware. Moreover, we evaluated the network's performance with varying user capacities under system cross-talk through a simulation. The results demonstrate that our network can support up to 64 users with key rates up to 1070~bps. Our network design benefits from qubit-based synchronization, enabling user access to QKD using simple, cost-effective devices.
        
        Our scheme shows promise for building quantum infrastructure for a large user base. In addition, benefitting from qubit-based synchronization, it naturally removes the effect of Raman scattering of synchronous light on quantum signals and is a promising new candidate for integrated quantum networks ~\cite{Wang2021,geng2021}, which sacrifice channel spacing to mitigate the impact of strong light on quantum signals. Our network can further reduce costs using a silicon-based transmitter chip~\cite{2016Ma-chip,2022Liu-review,2023Sax,2023Wei-chip,2023Du-chip}. Furthermore, in addition to secure key distribution, our scheme can apply to other cryptographic protocols,  such as secret sharing~\cite{2013wei,2023sheng-QS,sheng2023accessible}, quantum private query~\cite{liu2022decoy} \blk and digital signatures~\cite{2017Roberts,2022Yin-QS}.
         
        Finally, we conclude with some practical issues on the application of our proposed QAN.
     
        (1) Work mode of detectors: 
        in our network, free-running single-photon detectors are essential for qubit-based synchronization. Hence, we employ a free-running SNSPD in our demonstration. Moreover, we have the option to replace the detector with cost-effective free-running avalanche SPDs. Consequently, our method promises cost reduction for commercial QKD systems and enhances the attractiveness of our scheme.
        
        (2) Network users with different distances and system frequencies: 
        our demonstration focuses on network users with identical distances and frequencies. In practical networks, users have varying distances and system frequencies. We can compensate for time differences resulting from varying distances during the initial time synchronization, ensuring synchronization between each user and the receiver for successful time-division multiplexing. Additionally, our scheme is not suitable for users with arbitrarily different frequencies in principle. Fortunately, in traditional optical communication networks, typical user frequencies double. In these cases, each user needs only to periodically align with a predefined time window to facilitate clock recovery using our method. It is intriguing to investigate the feasibility of our method in these doubled-frequency scenarios.
       
        (3) Polarization feedback using qubit states: in a typical network, polarization feedback is required to maintain the low-error operation. We note that rapid polarization state feedback is not required in metropolitan area networks. As reported in the literature~\cite{2021avesani,2018Li.feedback}, polarization states in deployed optical fibers exhibit stable transmission for over 1-2 hours, and similar observations have been achieved in laboratory settings~\cite{2020Wei}. Therefore, once we align the polarization reference frame through calibration, subsequent QKD operations necessitate only low-speed and time-division polarization feedback in the network. In our recent work~\cite{2023Wei-chip}, we have demonstrated the effective use of qubit synchronization methods to recover the clock within optical fibers up to 150 km, enabling efficient polarization feedback. This method can extend to network settings, with the key distinction being the need to distinguish signals originating from different users and correspondingly apply time-division feedback.

	    \section*{Acknowledgment}
    	This study was supported by the National Natural Science Foundation of China (Nos. 62171144 and 11905065), Guangxi Science Foundation (Nos.2021GXNSFAA220011 and 2021AC19384), Open Fund of IPOC (BUPT) (No. IPOC2021A02), and Innovation Project of Guangxi Graduate Education (No.YCSW2022040).

	    \appendix 
    	\section{Time synchronization method}\label{Appendix_A}
\begin{figure}[htp]
    		\centering
    		\includegraphics[width=0.8\linewidth]{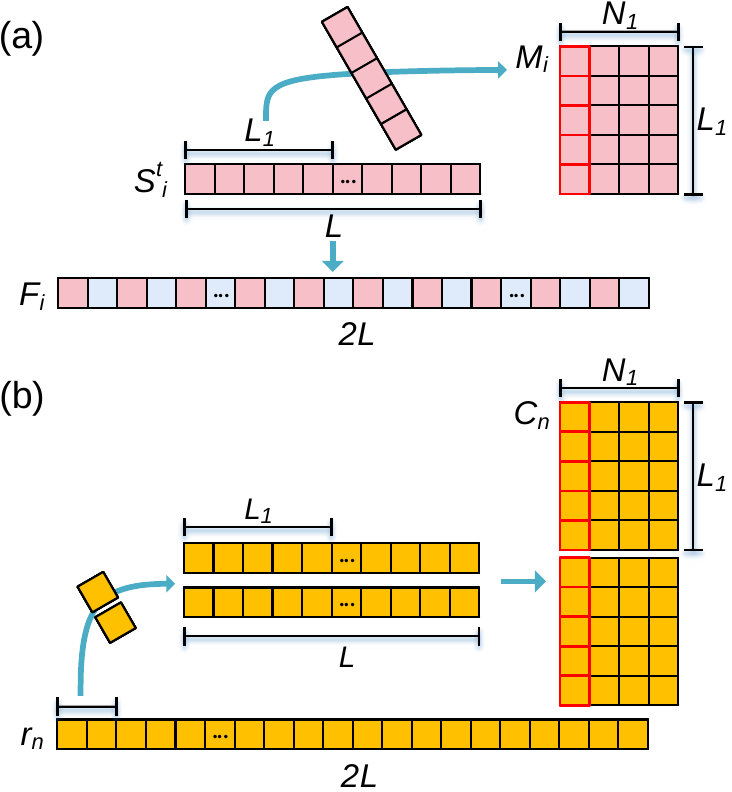}
    		\caption{(a) Synchronization string $S_{i}^{t}$ of length $L$ is reconstructed into the matrix $M_{i}$, where $L_{1}$ is the length of the small period and $N_{1}$ is the number of periods. (b) Received qubit frame $r_n$ of length $2L$ comes from a detection slot. The string is divided into two strings: one containing synchronous qubits and the other containing random qubits. Two strings are reshaped into two matrices $C_{n}$ and subjected to the cross-correlation operation with the public synchronization matrix $M_{i}$.}
    		\label{fig_identify}
    	\end{figure}

    	Here, we describe qubit-based synchronization in the QAN. Transmitter-$i$ performs periodic correlation encoding on the synchronization string $S_{i}^{t}$, splits it into single bits, and periodically embeds each bit in a random string to form a bit frame $F_{i}$, as shown in Figure~\ref{fig_identify}(a). We define $L$ as the length of the synchronization string, where $L_1$ is the length of the small period and $N_1$ is the number of periodic peaks. The length of a bit frame is $2L$, that is, each synchronization bit is followed by a random bit. The number of random bits can be adjusted to increase the effective secure key rate. For example, each synchronization bit connects $M$ random bits to form a bit frame of length $(M+1)L$. The frame is used to modulate a sequence of quantum signals based on the transmitter's local clock $\tau_{i}$, where the synchronization string is reused.
    	
    	Each transmitter sends qubits frame by frame. These qubits are sent to the receiver via a lossy channel. The receiver performs time synchronization based on the measured qubits. For transmitter-$i$, the receiver must determine the expected time of arrival $t_{i,a}^{e}$, which can be expressed as
    	
    	\begin{equation}
    	\begin{aligned}
    	t_{i, a}^{e} & = t_{i, 0}+n_{i, a} \tau^{R}+\epsilon_{i, a}, \quad n_{i, a} \in \mathbb{N},
    	\end{aligned}
    	\end{equation}
    	where $n_{i, a}$ is the order number of the qubit in the qubit string sent by transmitter-$i$, $a$ denotes the $a$-th detection, and $t_{i, 0}$ is the absolute time offset between transmitter-$i$ and the receiver. $n_{i, a}$ is a variable that satisfies the normal distribution, with expected value zero and variance $\sigma^{2}$.
    	
    	We first describe how the receiver recovers the clock $\tau^{R}$ from the measured signals. The clock is initially estimated based on the time-of-arrival signals by applying a fast Fourier transform. We sample the time of the measured signal at a sampling rate of $4/\tau_{i}$. The number of samples $N_f$ used to perform the Fourier transform is set to $10^6$. Then, the least trimmed square algorithm is applied to these sampled signals to obtain a more precise clock satisfying
    	\begin{equation}
    	\begin{aligned}
    	\frac{1}{D} \sum_{b=1}^{D}\left|\mathcal{E}_{a}^{I}(b)\right|^{2} \simeq \sigma^{2},
    	\end{aligned}
    	\end{equation}
    	where $\mathcal{E}_{a}=t_{a}^{m}-t_{a}^{e}$ denotes the error between the measured and expected times of arrival for all transmitter signals, and $t_{a}^{m}$ is the measured time-of-arrival of signals with $a\ge 1$. $\mathcal{E}_{a}^{I}(b)=\mathcal{E}_{a+b}-\mathcal{E}_{a}$ indicates the time interval error between two different detection events $a$ and $a+b$.The detection time slot $T_n$ in the clock window $\tau^{R}$ can be determined by 
    	\begin{equation}
    	\begin{aligned}
    	T_n = mod_{\tau^{R}}(t_{a}^{m}).
    	\end{aligned}
    	\end{equation}
    	The arrival time $t_{n, a}^{m}$ of the different transmitters’ detection events can be correctly divided from $t_{a}^{m}$ based on the recovered clock $\tau^{R}$. Owing to channel loss and background noise, the demarcated slots may deviate from the slots pre-designated by the calibration process. Therefore, the receiver must identify the transmitter corresponding to the time slot. 
    	
    	To determine which time slot transmitter-$i$ corresponds to and to recover the time offset $t_{i,0}$, the receiver calculates the correlation between the received frame and the public synchronization string $S_{i}^{t}$. The synchronization string is encoded on the $z$-basis and we label the quadrature-encoded qubits as $+1$ and $-1$. In Figure~\ref{fig_identify}(a), to reduce the complexity of the cross-correlation algorithm, the string $S_{i}^{t}$ with periodic correlation is converted into the matrix $M_i$, where $L_1$ is the length of the small period and $N_1$ is the number of periodic peaks.  
    	
    	We extract the received qubit frame from each detection time slot.  As shown in Figure~\ref{fig_identify}(b), the receiver extracts a received qubit frame $r_n$ of length $2L$, in which the qubits not detected are recorded as $0$, as are those clicked on the $x$-basis detector. The qubits that click the $z$-basis detector are assigned +$1$ and -$1$, consistent with the encoding definition of the synchronization string. Based on the structure of the bit frame sent by the transmitter, the received qubit frame $r_n$ can be divided into two strings, one for key generation and the other for synchronization. The two strings are reconstructed into two matrices $C_n$ following the rule of matrix $M_i$. By exploiting the correlation between the first column of matrices $C_n$ and the public matrix $M_i$, the maximal correlation helps identify the synchronization qubit string and the corresponding time slot. Finally, according to the maximum correlation position between the received matrix $C_i$ and the public matrix $M_i$, the time offset $t_{i,0}$ between the received frame and sent by transmitter-$i$ can be recovered.
    	 
    	\section{Practical issues for time synchronization method}\label{Appendix_B}
    		
    	In this section, we discuss with some practical issues on the application of the time synchronization.
    		
    	(1) Synchronization using classical communication: we note that while QKD necessitates classical communication for postprocessing, the latter cannot be directly utilized for QKD synchronization. The reasons are well discussed in previous studies such as Refs.~\cite{2020Calderaro,2023Berra}. Here, we give a briefly summarizations: (a) Classical communication systems typically utilize self-synchronizing codes for synchronization. Hence, an external synchronization system is not readily available to extend to QKD synchronization. (b) Employing classical systems for QKD synchronization necessitates a physical connection between ``classical" and ``quantum" hardware, alongside appropriate modification of classical transceivers. (c) QKD implementation does not demand synchronous classical postprocessing in relation to quantum communication and classical communication often operates at higher clock rates compared to QKD. Therefore, additional upgradation is needed to employ classical communication for quantum communication synchronization.
    		
    	(2) Requirements for time resolution of detector: establishing a precise model to analyze the requirement for the detector's time resolution is highly complex due to the involvement of numerous practical system parameters, including the pulse width and time jitter of light,  intrinsic error rate of the optical system, dark count of the detector, among others. In this context, we only employ simplified time-jitter models for a qualitative analysis.
    		
    	Here, we assume an ideal light pulse and a Gaussian distribution of arrival light due to time jitter. The probability of light arriving within a time slot can be calculated as
    	\begin{equation}
    	    \begin{aligned}
            P = \int_{-\frac{1}{2f}}^{\frac{1}{2f}} \frac{1}{\sqrt{2\pi} \sigma} \cdot e^{-\frac{t^2}{2\sigma^2}} dt,
    		\end{aligned}
    		\end{equation}
    		where ${f}$ is the system repetition frequency, $\sigma = \frac{T_{\text{FWHM}}}{2\sqrt{2\ln2}}$ represents the standard deviation, and $T_{\text{FWHM}}$ is the half-maximum full width of the time distribution caused by the time jitter of the detector. Assuming a zero error rate within the time slot and 50\% outside, we can calculate the total QBER caused by time jitter as
    		\begin{equation}
    		\begin{aligned} E = \frac{1-P}{2} \quad  	\end{aligned}
    	\end{equation}
    		
    	Combining Eqs. B1 and B2 allows us to determine the minimum time resolution. For a quantitative view, we consider a specific case where $f = 10$ GHz and the threshold $E \leq 11\%$ (a threshold value for secure key distribution), and we obtain the time resolution of the detector as $T_{\text{FWHM}} \leq 80$ ps. In practice, because of the presence of background and other error contributions, the required time resolution should be larger than the ideal case.
    
    	(3) Minimum number of counts of synchronization signals: we consider the case of zero error and no dark counts in the optical system and detector. Our scheme inherits the features of qubit-based synchronization, functioning effectively as long as the received signal's count can achieve a high signal-to-noise (SNR) ratio or low QBER. According to the formula provided in Ref.~\cite{,2020Calderaro}, the SNR can be quantified by $\Delta \approx \sqrt{L\eta}$, where $L$ is the length of the synchronization bit string, and $\eta$ is the total transmittance for each user. Furthermore, $\Delta \geq 10$ if a successful time recovery is performed. Thus, the minimum count for synchronization signals should be greater than 100. In practice,
    	the presence of background and misalignment between the
    	transmitter and the receiver enhance the requirement of minimum number of counts. Furthermore, we can determine the minimum length of the $L$ using $L\approx 100/\eta$.
    			
        \section{Channel model with cross-talk}\label{Appendix_C}
        
    	We simulate the secure key rate per user based on the measured cross-talk result. Intuitively, cross-talk leads to an increase in the gain and error rate. The gain and quantum bit error rate are similar to those of the model considered by Fr\"{o}hlich et al.~\cite{2013.Frohlich.Nature}, except for excluding the after-pulse of the detector. Here, we use an SNSPD to detect quantum signals. The SNSPD is also possibly subject to after-pulsing that  arises not from the SNSPD itself, but rather from the electronic readout module~\cite{Fujiwara2011,Marsili2012,Burenkov2013}. Marsili et al.~\cite{Marsili2012} showed that this after-pulsing
    	can be eliminated with appropriate treatment. Thus, the gain and QBER for intensity $k \in \{\mu,\nu\}$ in $z$-basis are given by
    	\begin{equation}
    	\begin{aligned}
    	Q_{z,k}&=k\eta (1+\frac{p_{T}}{N_c-1}(n-1))+p_{dc},\\
    	E_{z,k}&=\frac{k \eta(p_{opt}+\frac{1}{2}\frac{p_{T}}{N_c-1}(n-1))+\frac{p_{dc}}{2} }{Q_{z,k}},
    	\label{QBER}
    	\end{aligned}  
    	\end{equation}
    	where $N_c$ is the network capacity, $n$ is the number of active users, $p_{dc}$ is the dark count rate, and $p_{opt}$ is the optical error owing to misalignment. The overtransmittance $\eta$ of the system is given by $\eta=\eta_{ch}\eta_{spl}\eta_{r}$, which includes the channel, splitter transmittance, and the entire receiver transmittance. The channel transmittance is $\eta_{ch}=10^{-\alpha L/10}$, where $L$ is the fiber length and $\alpha$ is the loss coefficient of standard fiber. In practice, each user generally has different channel lengths corresponding to different channel transmittance $\eta_{ch}$. The splitter transmittance $\eta_{spl}$ can be calculated using the splitter parameters of a typical QAN experiment~\cite{2013.Frohlich.Nature}. The splitting ratio of a $1\times64$ splitter is $19.5$ dB. The count increase rate $p_{T}$ can be extracted from the cross-talk data shown in Figure~\ref{fig_crosstalk_simulation}(a). The specific values of these parameters are listed in Table~\ref{experiment_parameter}. Finally, the QBER in the $z$-basis is given by
    	\begin{equation}
    	\begin{aligned}
    	e_{z}&=\frac{P_{\mu}E_{z,\mu}+P_{\nu}E_{z,\nu}}{P_{\mu}Q_{z,\mu}+P_{\nu}Q_{z,\nu}}.
    	\end{aligned}
    	\label{QBER}  
    	\end{equation} 
    	
    	\begin{table}[H]
    		\centering
            \caption{Practical parameters for numerical simulations, including the optical misalignment $p_{opt}$, dark count rate $p_{dc}$ and total receiver transmittance $\eta_{r}$. The count increase rate caused by cross-talk is represented by $p_{T}$. $\alpha$ denotes the loss coefficient of standard fiber.}\label{experiment_parameter}
            \
    	    \doublerulesep 0.1pt \tabcolsep 10pt
    			\begin{tabular}{ccccc}
    				\hline
    				$p_{opt}$ & $\eta_{r}$ & $p_{dc}$ & $p_{T}$ & $\alpha$\\ 
    				\hline
    				$1\%$ & $38.8\%$ & $6\times10^{-8}$ & $0.98\%$ & $0.2$ dB/km\\ 
    				\hline
    			\end{tabular}
    	\end{table}
	    
     	\bibliography{QAN}
        \end{document}